\documentclass[aps,twocolumn,groupedaddress,superscriptaddress,showpacs,prl,floatfix]{revtex4}

\usepackage{amsmath}
\usepackage{graphicx}
\usepackage{dcolumn}
\usepackage{bm}
\usepackage{epsfig}
\usepackage{dsfont,psfrag}
\usepackage{color}
\usepackage{ifthen}

\newcommand{\ket}[1]{{| #1\rangle}}
\newcommand{\sgn}{\operatorname{sgn}}
\newcommand{\I}{\mathrm{i}}
\newcommand{\E}{\mathrm{e}}
\def\refeq#1{(\ref{#1})}
\begin{document}
  \title{Transferring entangled states through spin chains by boundary-state multiplets}
\author{Peter Lorenz}
\author{Joachim Stolze}
\email{joachim.stolze@tu-dortmund.de}
\affiliation{Technische Universit\"at Dortmund, Fakult\"at Physik, D-44221 Dortmund, Germany}

\date{\today}
\begin{abstract}
Quantum spin chains may be used to transfer quantum states between
elements of a quantum information processing device. A scheme
discovered recently \cite{BFR+12} was shown to have favorable transfer
properties for single-qubit states even in the presence of built-in
static disorder caused by manufacturing errors. We extend that scheme
in a way suggested already in \cite{BFR+12} and study the transfer of
the four Bell states which form a 
maximally entangled basis in the two-qubit Hilbert space. We show that
perfect transfer of all four Bell states separately and of arbitrary linear combinations 
 may be achieved for chains
with hundreds of spins. For simplicity we restrict ourselves to
systems without disorder.
\end{abstract}
\maketitle

Quantum information processing \cite{NC01} relies on a number of key
elements of technology, among them quantum bits and quantum
gates. Since any  quantum computer will contain a large number of
different quantum gates and registers, information must be transferred
between these elements of the computer. One possibility for that
information transfer is offered by quantum spin chains, linear arrays
of suitably coupled qubits. Research on quantum information transfer
by spin chains started roughly a decade ago \cite{Bos02} and quickly
developed into an active field with many contributors (see, for
example, the reviews in \cite{NJ13}). However, most of the research up
to now has focused on the transfer of single-qubit states, although the
handling of entangled multi-qubit states is of primary importance in
all known algorithms of quantum information processing.
In this Brief Report we show how a natural extension of a single-qubit
state transfer protocol \cite{BFR+12} can be used to achieve
high-fidelity transmission of arbitrary pure two-qubit states along
spin chains with up to hundreds of sites.

Spin chains for quantum information transfer mostly fall into one of
two classes distinguished by the degree of ``engineering'', or
fine-tuning, of the nearest-neighbor couplings along the
chain. Perfect state transfer (PST) may be achieved if all transition
frequencies generated by the spin chain Hamiltonian are commensurate
and hence the time evolution of arbitrary initial states becomes
periodic. In combination with spatial symmetry this enables perfect
``mirroring'' of initial states located at one end of the chain
\cite{CDEL04,my44,YB04,SLS+05,note2}. To achieve this goal, all
nearest-neighbor couplings must be tuned to specific values, hence
this class of chains may be called ``fully engineered''. A much
simpler route to good (but not perfect) state transfer is opened by
modifying only the boundary couplings affecting the very first and
last spins of the chain, respectively,  leaving all other couplings at
one and the same value \cite{WLK+05,BACx10,ZO11}. In the limit of weak
boundary couplings the system then possesses nearly degenerate
(symmetric and antisymmetric) eigenstates concentrated on the boundary
spins and the dynamics of these states may be exploited for the
transfer of quantum information. That class of systems may be called
boundary-dominated or optimizable, since different choices of the
boundary couplings may be used to adjust fidelity and speed of the
state transfer. The approach suggested in \cite{BFR+12} is an
interesting hybrid between the fully-engineered and boundary-dominated
schemes. As will be explained in more detail below, the temporal
structure (commensurate energy spectrum leading to perfect
periodicity) and the spatial structure (boundary-localized states
insensitive to perturbations originating in the interior of the chain)
can be optimized at the same time.

The system under consideration is a nearest-neighbor coupled spin-1/2
XX chain with spatially symmetric couplings:
\begin{equation}
  \label{eq:1}
  H=\frac 12 \sum_{i=1}^{N-1} J_i
 (\sigma_i^x \sigma_{i+1}^x +\sigma_i^y \sigma_{i+1}^y); \;
 J_i=J_{N-i} > 0 ,
\end{equation}
where $\sigma_i^{x,y}$ are Pauli matrices. The total $z$ spin component
is conserved, $[H, \sum_i \sigma_i^z]=0$, hence subspaces of fixed
total $z$ spin component can be treated separately. 
By the Jordan-Wigner transformation
\cite{JW28},
the spin chain (\ref{eq:1})
can be mapped to non-interacting lattice fermions with
nearest-neighbor hopping, where an up spin maps to a (spinless) fermion, while a
down spin maps to an empty site. Consequently, the Hamiltonian is
diagonalized once the single-particle eigenstates $|{\nu}\rangle$ and the
corresponding eigenvalues $\varepsilon_{\nu}$ are known. The
single-particle Hamiltonian is a symmetric tridiagonal matrix with
zeros on the diagonal and the couplings $J_i$ next to the diagonal. The
special form of that matrix determines several properties of the
spectrum. The  $\varepsilon_{\nu}$ come in pairs
$\pm|\varepsilon_{\nu}|$, the corresponding eigenvectors being related
by a sign reversal of every other component. Successive eigenvectors
(as ordered by energy) are alternatingly even and odd under spatial
reflection. This property makes perfect state transfer possible if the
$\varepsilon_{\nu}$ are commensurate (see, for example, \cite{my64}
for details). A prominent example \cite{note3} is given by 
$J_i=\sqrt{i(N-i)}$,
leading to an equidistant ladder of $\varepsilon_{\nu}$
values. However, it is even possible to prescribe a set of eigenvalues
$\varepsilon_{\nu}$ and find the corresponding couplings $J_i$ by
solving a structured inverse eigenvalue
problem \cite{Gla04,CG05}. There are several algorithms for solving
inverse eigenvalue problems; here we follow \cite{BFR+12} in using an
algorithm by de Boor and Golub \cite{BG78}. A PST chain with
particularly benign properties \cite{BFR+12} is defined by the
``inverted quadratic spectrum''
\begin{equation}
  \label{eq:2}
  \varepsilon_{\nu} = \nu(N-1-|\nu|); \;
  \nu=-\frac{N-1}2,...,\frac{N-1}2 .
\end{equation}
Note that the differences between successive  $\varepsilon_{\nu}$ are
largest close to the center of the spectrum, i.e. to $\nu=0$, for odd
$N$, the case on which we concentrate from now on. We note in passing
that the spectrum (\ref{eq:2}) superficially resembles the
cosine-shaped spectrum of the homogeneous ($J_i \equiv J$) XX chain;
the couplings $J_i$ corresponding to  (\ref{eq:2}), however, are roughly
constant only in the central region of the chain and oscillate
significantly towards the boundaries \cite{BFR+12}.

In contrast to the fully-engineered approach exemplified by the
spectrum (\ref{eq:2}), the boundary-dominated approach to quantum
state transfer employs a simple pattern of couplings, 
\begin{equation}
  \label{eq:3}
  J_1=J_{N-1}= \alpha J; \; J_i=J \mbox{ for } i \ne 1, N-1,
\end{equation}
where $\alpha < 1$ is an adjustable parameter. For small $\alpha$ a
perturbation calculation shows that there are three closely spaced
energy eigenvalues close to the center of the spectrum ($N$ is odd); the
corresponding eigenvectors are dominantly localized on the boundary sites and
thus can be used to transfer information back and forth between the
ends of the chain. Clearly, as $\alpha$ gets smaller, the influence
of the interior spins decreases and the fidelity of the state transfer
increases, but so does the transfer time which is inversely
proportional to the energy splitting between the states of the
dominant triplet.

The combination of PST and boundary-dominated state transfer
\cite{BFR+12} rests  on the following key observation:
A PST chain with spectrum (\ref{eq:2}) may be equipped with a triplet
of closely spaced energies by simply ``contracting'' it towards the
center:
\begin{equation}
  \label{eq:4}
  \varepsilon_{\nu}^{\prime} =  \varepsilon_{\nu}- (N-3) \sgn
  \varepsilon_{\nu}
\end{equation}
($ \sgn 0$ is to be interpreted as zero).
The new spectrum $ \varepsilon_{\nu}^{\prime}$ is still commensurate,
ensuring PST, but the triplet of states with energies close to zero
are strongly boundary-dominated, as with the couplings (\ref{eq:3}),
for small $\alpha$. As a bonus, the temporal behavior of the
probability to collect the transmitted state at the receiving end of
the chain changes from an extremely spiky shape with a needle-like
maximum to a single broad and smooth maximum, thus making it much
easier to measure the transmitted state at the right instant. This
feature can be intuitively understood from the fact that the states
dominating the transfer involve only small energy differences,
i.e. long time scales.

The state transfer scheme just discussed can be extended quite
naturally, as already  suggested in \cite{BFR+12}: collecting additional
closely spaced eigenvalues in the center of the spectrum generates
boundary states on more sites. A contraction like (\ref{eq:4}) with
$(N-5)$ in place of $(N-3)$ and acting on all levels except the
previously created triplet generates a quintuplet of levels with unit
spacing. A look at the coupling constant distribution shows that the
region of nearly constant $J_i$ around $i= \frac{N-1}2$ shrinks with
every contraction step, giving way to oscillatory behavior near the
ends of the chain.

\begin{figure}[h]
\includegraphics[width=\columnwidth,clip]{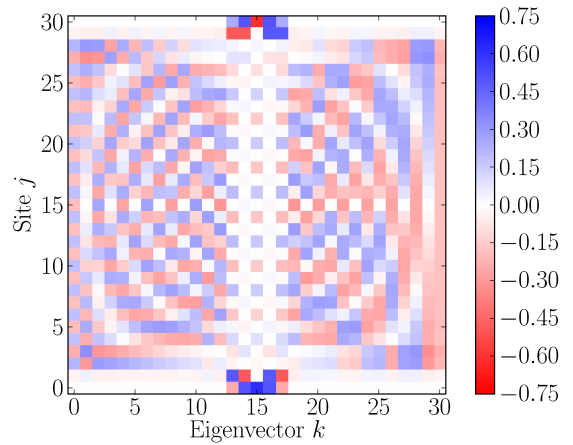}
\caption{(Color online) Eigenvectors for a chain with a quintuplet of
  closely spaced energies, $N=31$. The states of the quintuplet are
  basically localized on the two pairs of sites close to the ends of
  the chain.}
\label{eins}
\end{figure}

The spatial structure of the energy eigenstates is shown in
Fig. \ref{eins}. The states of the quintuplet clearly show large
weights on the two first and last sites of the chain, respectively,
while all other states have negligible weight there. 

\begin{figure}[h]
\includegraphics[width=\columnwidth,clip]{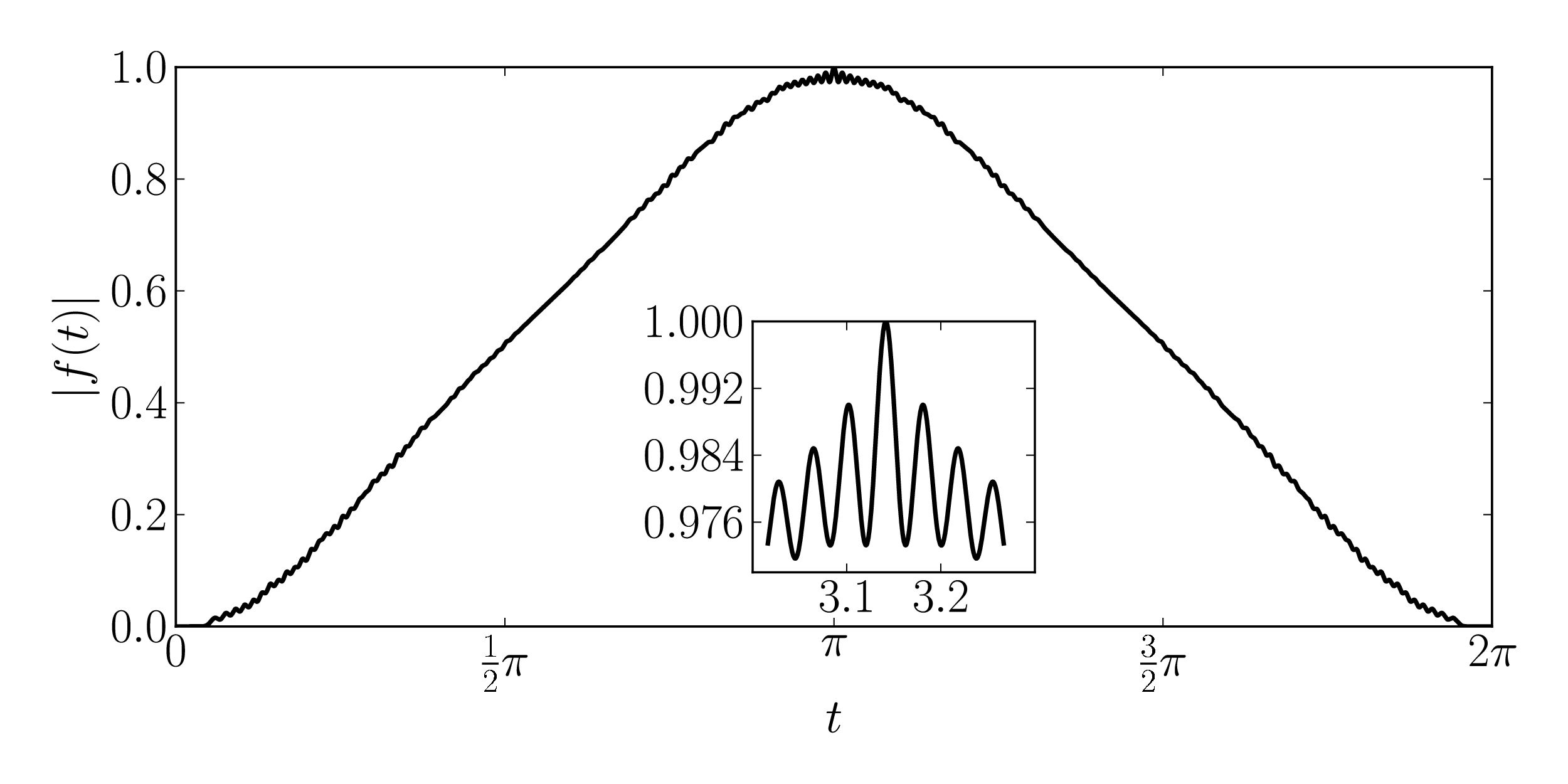}
\caption{Fidelity of transmission \refeq{eq:6} 
of the Bell states $| \psi_1 \rangle_{\pm}$  for a
  chain with a quintuplet of closely spaced energies, $N=31$. The
  inset shows the region close to the perfect transfer time, $t=\pi$.}
\label{zwei}
\end{figure}

The transfer properties of the system were studied for the set of Bell
states, which form a maximally entangled basis in the two-spin Hilbert
space:
\begin{equation}
  \label{eq:5}
  |\psi_1 \rangle_{\pm} = \frac 1{\sqrt 2} (|\uparrow \downarrow\rangle
  \pm |\downarrow \uparrow \rangle) ; \;
 |\psi_2 \rangle_{\pm} = \frac 1{\sqrt 2} (|\uparrow \uparrow\rangle
  \pm |\downarrow \downarrow \rangle) .
\end{equation}
In terms of Jordan-Wigner fermions, $  |\psi_1 \rangle_{\pm} $ belong
to the one-particle subspace of the full chain Hamiltonian
\refeq{eq:1}, while  $|\psi_2 \rangle_{\pm}$ contain two-particle
components and zero-particle components, the latter having trivial
dynamics. We denote by $| \phi_i\rangle$  the state with the first two
spins in one of the states \refeq{eq:5}, with all other spins down, and by
$| \phi_f\rangle$ the spatial mirror image of  $|
\phi_i\rangle$. Then, a convenient measure for the fidelity of
transmission can be defined by
\begin{equation}
  \label{eq:6}
  |f(t)| = |\langle\phi_f |\E^{-\I Ht}|\phi_i \rangle|.
\end{equation}
Fig. \ref{zwei} shows $  |f(t)|$ for the Bell states   $|\psi_1
\rangle_{\pm}$ with the $N=31$ chain discussed above. ($|f(t)|$ is
identical for the two states.)

\begin{figure}[h]
\includegraphics[width=\columnwidth,clip]{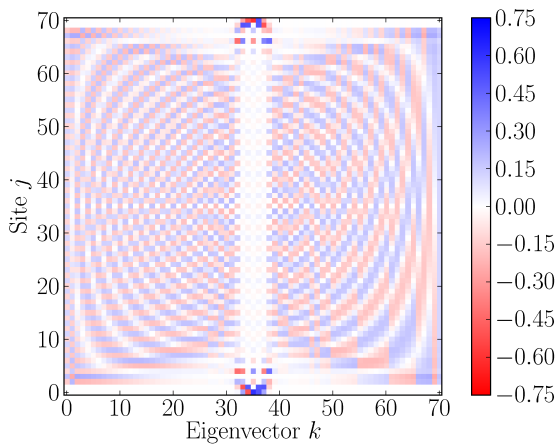}
\caption{(Color online) Eigenvectors for a
  chain with a quintuplet of closely spaced energies, $N=71$, and an
  additional contraction by $\Delta=60$. Note that seven states in the
  center of the energy spectrum are concentrated near the boundaries,
  while all other states extend through the whole system.}
\label{vier_neu}
\end{figure}

\begin{figure}[h]
\includegraphics[width=\columnwidth,clip]{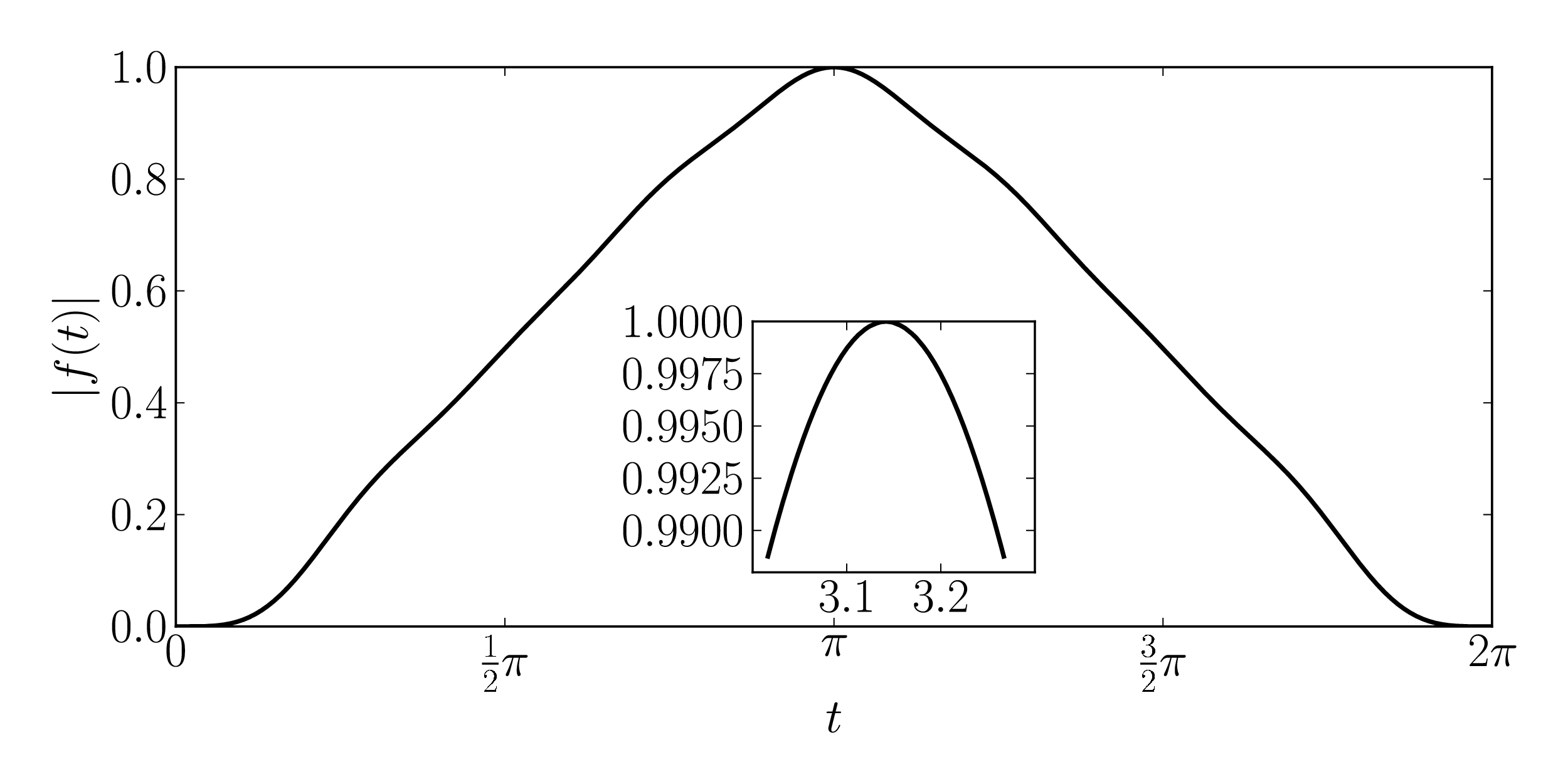}
\caption{Fidelity of transmission \refeq{eq:6} 
of the Bell states $| \psi_1 \rangle_{\pm}$  for a
  chain with a quintuplet of closely spaced energies, $N=71$, and an
  additional contraction by $\Delta=60$. The
  inset shows the region close to the perfect transfer time, $t=\pi$.}
\label{drei}
\end{figure}

Perfect transfer at $t=\pi$ is possible, but small high-frequency
oscillations are visible. They are due to the admixture of
higher-energy states. The oscillations become stronger for longer
chains, making PST extremely difficult beyond $N \approx 60$, because
the fidelity maximum at $t=\pi$ becomes extremely narrow. This can be
remedied by another contraction of the spectrum. Subtracting $\Delta
\sgn \varepsilon_{\nu}$, with $\Delta=60$, from all energies
outside the quintuplet leads to the spectrum $ \varepsilon_{\nu} =0,
\pm 1, \pm 2, \pm 7, \pm 70,...$ for the $N=71$ chain, which displays
a smooth fidelity vs. time curve, see Fig. \ref{drei}. The
eigenvectors for $N=71$ and $\Delta=60$ are shown in
Fig. \ref{vier_neu}.
Comparison to Fig. \ref{eins} shows that the quintuplet of
boundary-dominated states is on its way to develop into a
septuplet. Furthermore it should be noted that the quintuplet contains
almost the complete weight of states localized on the first two
lattice sites. This is definitely different for $\Delta=0$ (not shown
here), where the position eigenstate on site 2 contains significant
weight from energy eigenstates close to the boundaries of the
spectrum. This is what causes small high-frequency oscillations in the
transfer fidelity of the single-qubit state initially located at site
2, and consequently, of the Bell states $| \psi_1 \rangle_{\pm}$
as shown, for example, in Fig. \ref{zwei} for $N=31$.

%
\begin{figure}[h]
\includegraphics[width=\columnwidth,clip]{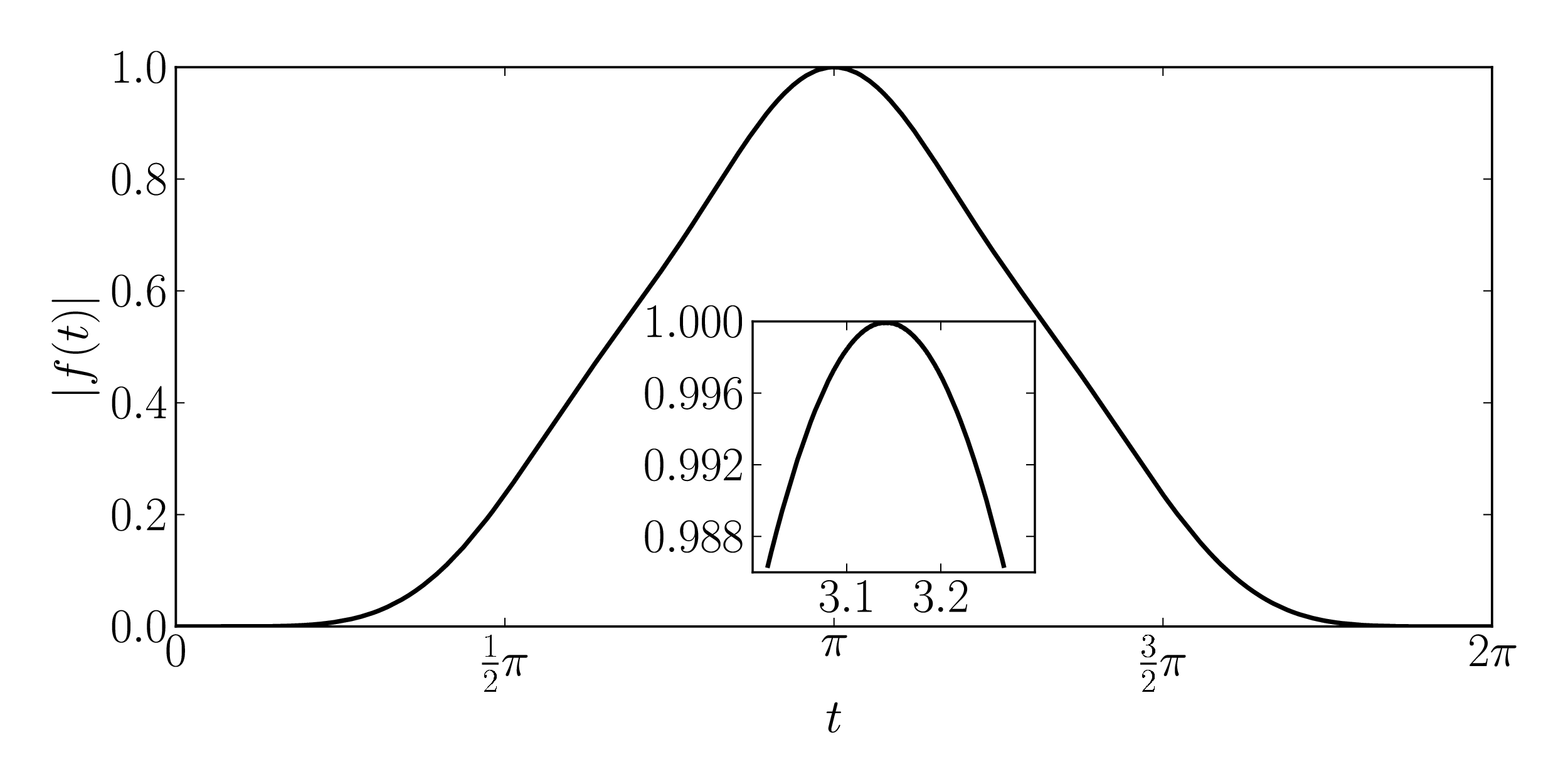}
\caption{Fidelity of transmission \refeq{eq:6}
of the Bell states $| \psi_1 \rangle_{\pm}$ for a
  chain with a nonuplet of closely spaced energies, $N=321$. The
  inset shows the region close to the perfect transfer time, $t=\pi$.}
\label{vier}
\end{figure}

The recipe just demonstrated for $N=71$ is also applicable for longer
chains. For $N \approx 200$ the optimal contraction parameter turns
out to be $\Delta=N-7$, meaning that the quintuplet of levels with
unit spacing has grown into a septuplet. Continuation of the process
generates a nonuplet of states which, for example, leads to decent
behavior for $N=321$, as shown in Fig. \ref{vier}.

\begin{figure}[h]
\includegraphics[width=\columnwidth,clip]{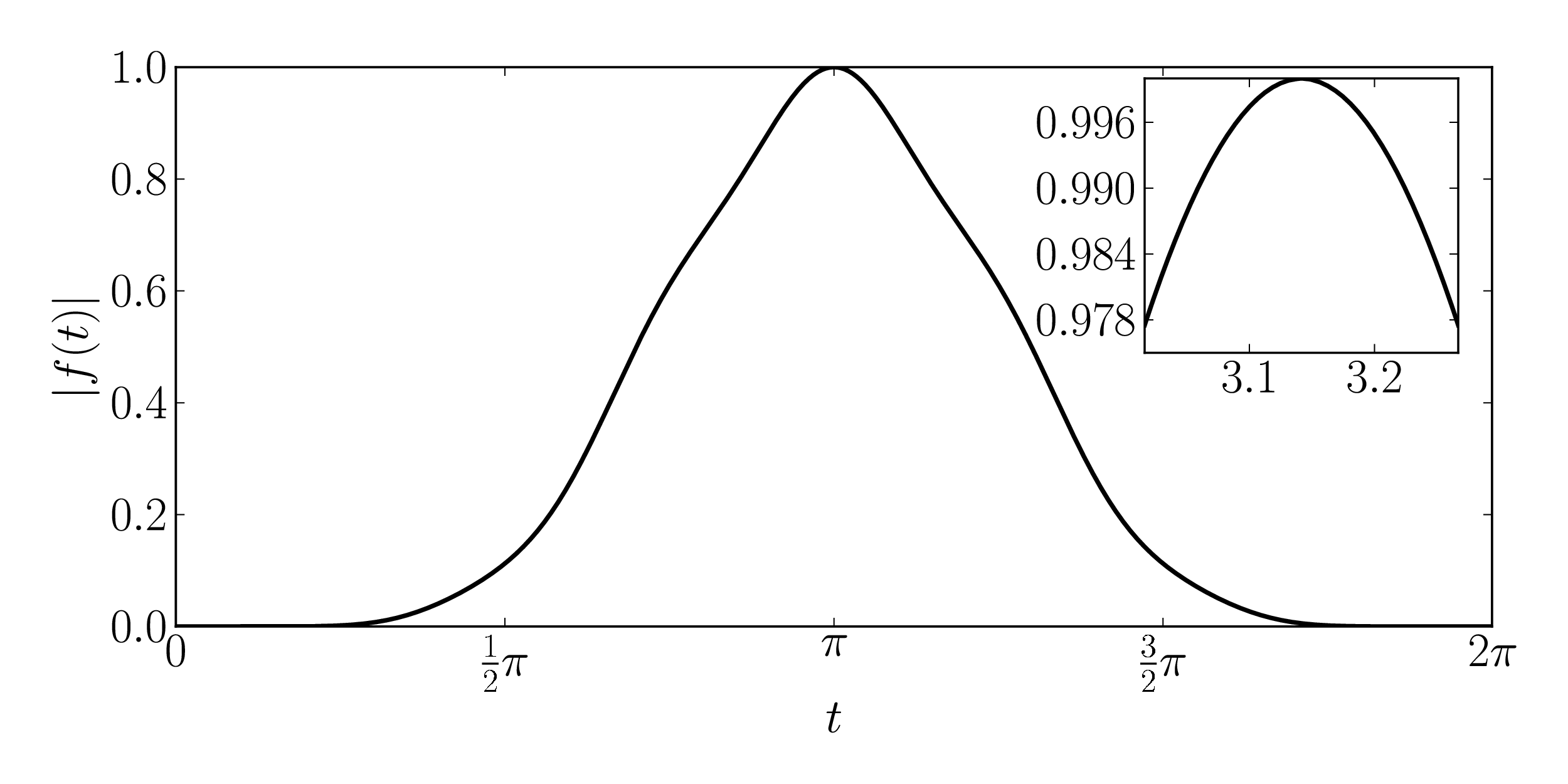}
\caption{Fidelity of transmission \refeq{eq:6} 
of the Bell states $| \psi_2 \rangle_{\pm}$  for a
  chain with a quintuplet of closely spaced energies, $N=71$, and an
  additional contraction by $\Delta=60$. The
  inset shows the region close to the perfect transfer time, $t=\pi$.}
\label{fuenf}
\end{figure}

The Bell states $| \psi_2 \rangle_{\pm}$ involve the dynamics of
two-particle states. Since the vacuum component of those states is
trivial, the transfer fidelities of  $| \psi_2 \rangle_{\pm}$ are
equal. It turns out that the fidelity of  $| \psi_2 \rangle_{\pm}$ is
slightly more delicate than that of $| \psi_1 \rangle_{\pm}$, with
narrower maxima and stronger oscillations. Nevertheless, the
configuration with $N=71$ and additional contraction by $\Delta=60$
yields similar excellent transfer properties for states $| \psi_1 \rangle_{\pm}$
(Fig. \ref{drei}) and  $| \psi_2 \rangle_{\pm}$ (Fig. \ref{fuenf}.)

Since all Bell states can be transferred perfectly, one is tempted to
conclude that PST is possible for arbitrary two-spin states, the Bell
states forming a basis in the two-spin Hilbert space. However, this is
not so, due to the presence of different phase factors \cite{note4}.
To discuss
those it is convenient to use a description in terms of Jordan-Wigner
fermions. 
Phase differences occur between states with different numbers of up
spins (Jordan-Wigner fermions). Two mechanisms are responsible for
these phase differences. Firstly, the perfect transfer of every up
spin is accompanied by a fixed phase shift, and secondly, sign changes
occur due to the statistics of the Jordan-Wigner fermions. 

The
phase $\varphi$ involved in the PST of single-spin (single-fermion)
states by mirroring \cite{CDEL04, my44, YB04, SLS+05} is defined by
\begin{equation}
  \label{eq:7}
  \E^{-\I H \tau} \ket{i} =\E^{\I \varphi} \ket{N+1-i}.
\end{equation}
Here, $\ket{i}$ denotes the state with a single up spin at site $i$
and down spins everywhere else; $\tau$ is the perfect transfer
time. The phase $\varphi$ does not depend on the site index $i$; for
the PST chain with $J_i=\sqrt{i(N-i)}$ \cite{CDEL04}
\begin{equation}
  \label{eq:8}
  \varphi= -\frac{\pi}2 (N-1).
\end{equation}
This can be derived from the analogy between the rotation of a spin-$S$
particle in a transverse magnetic field and the motion of a particle
along a chain with $2S+1$ sites, each site corresponding to an $S_z$
eigenstate, as discussed, for example, in \cite{CMJ05}. 
It turns out that (\ref{eq:8})
also holds for general PST chains of odd length $N$. This follows
from the general properties of the eigenvectors and
eigenvalues of the tridiagonal single-particle Hamiltonian matrix
described earlier. For even $N$ there are two cases which have to be
distinguished. Note that for all PST chains, differences between
neighboring single-particle energies must be odd numbers (in
appropriate units). Thus, for even $N$ the two smallest (in absolute
value) eigenvalues must be $\pm |l+\frac 12|$ with some integer $l$,
and  (\ref{eq:8}) changes to
\begin{equation}
  \label{eq:9}
  \varphi= -\frac{\pi}2 (N-1) -\pi l.
\end{equation}
To conclude the discussion of the single-particle phase we note that
$\varphi$ may be adjusted by adding a constant magnetic field in $z$
direction, corresponding to a nonzero constant diagonal in the
Hamiltonian matrix.

A two-particle state with up spins at sites $i$ and $j>i$ is
equivalent to a two-fermion Fock state:
\begin{equation}
  \label{eq:10}
  \ket{i,j}= c_i^{\dag}  c_j^{\dag} \ket 0
\end{equation}
where $\ket 0$ is the vacuum (all spins down) state. The PST property
(\ref{eq:7}) in combination with Fermi statistics, determines the time
evolution of that state:
\begin{eqnarray}
  \label{eq:11}
    \E^{-\I H \tau} \ket{i,j} &=& \E^{-\I H \tau} c_i^{\dag}  c_j^{\dag}
    \ket 0\\ \nonumber &=&  \E^{2 \I \varphi} c_{N+1-i}^{\dag}  c_{N+1-j}^{\dag} \ket
    0\\ \nonumber
&=& -  \E^{2 \I \varphi} c_{N+1-j}^{\dag}  c_{N+1-i}^{\dag} \ket 0\\ \nonumber &=&
\E^{\I (2 \varphi + \pi)} \ket{N+1-j, N+1-i}.
\end{eqnarray}
The generalization to a state with $n$ fermions is straightforward;
the phase picked up during PST then is
\begin{equation}
  \label{eq:12}
  \phi_n = n \varphi + n(n-1) \frac{\pi}2.
\end{equation}
For the Bell states this means that $\ket{\psi_1}_{\pm}$
and $\ket{\psi_2}_{\pm}$ are transferred with different phase
factors, even if $\varphi=0 (\mbox{mod }2\pi)$ can be achieved by
choosing $N=4k+1$, for example. Furthermore, if the spins at sites 3 through $N$ are not
initialized to the down state, particle-number dependent phase factors
will mix up the transferred state.

However, if initialization is possible, there is a protocol that 
achieves PST with equal phases for all Bell states. The single-particle
phase should be adjusted to $\varphi=\frac{\pi}2$ (either by making
$N$ even or by an external field). Then $\phi_n$ is effectively zero
for even particle number $n$ and $\frac{\pi}2$ for odd $n$. The first
two sites then are initialized to an arbitrary superposition of the
four Bell states, while all other sites stay initialized to the down
state. Subsequently two controlled-$X$ gates are applied, with qubits
1 and 2 as control and (say) qubit 3 as target qubits: $ CX(1,3)
CX(2,3)$. Obviously this creates a superposition of states with only
even total particle numbers which can be transferred without additional
phase factors.

To summarize, we have extended the quantum information transfer scheme
given in \cite{BFR+12} from single-qubit states to arbitrary two-qubit
states by studying the transfer fidelity for all four Bell states for
states of up to hundreds of spins, in the absence of built-in
disorder. Phase factors causing unwanted interference between
Bell states containing even and odd numbers of up spins can be
compensated for.  Our approach follows the lines already suggested in
\cite{BFR+12}. The transfer of two-qubit states is achieved by
creating a quintuplet of closely spaced equidistant energy eigenvalues
in the center of the spectrum. The corresponding states suffice to
expand arbitrary two-qubit states. We have shown how additional
manipulations of the energy spectrum may be used to transfer Bell
states along progressively longer states.  These additional
modifications essentially create higher multiplets of energy
eigenstates which are concentrated on more boundary sites and may
serve to transfer also states of three or more qubits. The protocol
compensating for unwanted phase factors may also be naturally extended
to states involving three or more spins.

\bibliographystyle{apsrev}
\bibliography{/home/stolze/BIBLIO/jsbas_def,/home/stolze/BIBLIO/general,/home/stolze/BIBLIO/mypapers,/home/stolze/BIBLIO/qcomp,boundary_bell}

\begin{thebibliography}{10}
\bibitem{note2}
\bibinfo{note}{PST chains may be used to transfer states involving arbitrary
  numbers of qubits, however, the robustness of that process in the presence of
  perturbations has not been studied in detail yet.}
\bibitem{note3}
\bibinfo{note}{This system has been discovered and used in several branches of
  physics, see \cite{my64} for some examples}.
\bibitem{note4}
\bibinfo{note}{In fact, these phase differences have been employed earlier for
  constructing entangling gates \cite{YB04, CMJ05}.}
\end{thebibliography}

\end{document}